\documentstyle[12pt, iopfts]{ioplppt}

\begin{document}

\sloppy

\jl{2}

\title{Quantum Theory of High Harmonic Generation \\ 
via Above Threshold Ionization and 
Stimulated Recombination}


\author{M. Yu. Kuchiev\footnote{E-mail: kmy@newt.phys.unsw.edu.au}
 and V. N. Ostrovsky\footnote{
Permanent address: 
Institute of Physics, The University of St Petersburg,
198904 St.Petersburg, Russia; E-mail: Valentin.Ostrovsky@pobox.spbu.ru
}
}

\address{School of Physics, University of New South Wales
Sydney 2052, Australia}


\begin{abstract}
Fully quantum treatment explicitly presents the high harmonic 
generation as a three-stage process: above threshold
ionization (ATI) is followed by the continuum electron propagation 
in a laser field and subsequent stimulated 
recombination back into the initial state. The contributions
of all ATI channels add up coherently. All three stages of
the process are described by simple, mostly analytical
expressions. A very good quantitative agreement with 
the previous calculations on the harmonic generation by 
H$^-$ ion is demonstrated, thus supplementing the conceptual 
significance of the theory with its practical efficiency.
\end{abstract}



Under the influence of an intensive electromagnetic field an atom 
can emit electrons and photons. 
The number of photons absorbed from the field in the first process
generally exceeds the minimum necessary for ionization,
i.e. the photoelectrons are characterized by their distribution
over above threshold ionization (ATI) channels.
The photon production corresponds to the harmonics 
generation (HG) for the incident (monochromatic) laser radiation. 
Both ATI and HG are capable of populating the channels with 
remarkably high energy, as has recently been registered in 
experiments (see, e.g., L.'Huillier {\it et al}\/ 1991,
Macklin {\it et al}\/ 1993, Schafer {\it et al}\/ 1993) 
and tackled by the theory.

An idea that the two processes referred above are interrelated has
been articulated quite long ago. Since in the HG process
an active electron ends up in the initial bound state,
it is appealing to represent it as ionization followed by
recombination. This mechanism presumes a strong interaction
between the emitted electron and the core that is omitted
in the standard Keldysh (1964) model of multiphoton
ionization. The importance of this interaction for a 
variety of processes was first pointed out by Kuchiev (1987), 
who predicted several phenomena for which the electron-core 
interaction plays a crucial role. The related 
mechanism was named {\it ``atomic antenna''}.

Specifically for HG, the simple relation between this process 
and ATI was suggested by Eberly {\it et al}\/ (1989) but 
proved to be non-realistic, see below. 
The hybrid classi\-cal-quantum model due to Corkum (1993) casts HG 
as a three-step process: tunneling ionization
and subsequent propagation in the continuum is followed by 
recombination. This intuitive model has influenced many
research in experiment and theory. The simplicity
of the model is due to some 
drastic presumptions.
Usually it is emphasized that the intermediate electron 
propagation in the laser field is described by the 
Corkum (1993) model classically. Probably less
attention is paid to the fact that neither the tunneling 
ionization through the time-dependent barrier, nor 
the laser-stimulated recombination receive a genuine 
quantum treatment as well. Being successfully applied to 
the comparison with some experimental data, the 
model resorts to such a loosely defined free parameter as the 
'transverse spread of the electron wave function'. From 
the conceptual side the Corkum (1993) model
does not directly appeal to ATI process just because the
discrete ATI channels do not appear within the classical framework. 
The subsequent developments were based on more sophisticated 
theory and led to important advancements 
(Lewenstein {\it et al}\/ 1994, Becker {\it et al}\/ 1994),
but apparently abandoned the perspective to establish a quantitative 
relation between ATI and HG. The ATI characteristics
merely do not emerge in the papers devoted to HG theory,
with few exceptions (Zaretskii and Nersesov, 1996).

The principal objective of the present study is to derive a fully 
quantum formulation for the HG amplitude in terms of the ATI 
amplitude and amplitude of electron stimulated recombination (SR) 
in the laser field. Importantly, all the amplitudes are physical, 
i.e. no off-energy-shell entities appear. This circumstance adds to 
the conceptual appeal of the present theory its significance
as a true working tool. We successfully test its efficiency by
quantitative comparison with the benchmark calculations
by Becker {\it et al}\/ (1994) for HG by H$^-$ ion. 
In the broader perspective it should be emphasized that our
theoretical technique is directly applicable 
to other processes of current key interest, such as 
multiple ionization by laser radiation or photoelectron 
rescattering. 

The rate of the $N$-th harmonic generation is proportional
to $| d_N |^2$, where $d_N$ is the $N$-th Fourier component
of the transition dipole momentum $d(t)$ 
\begin{eqnarray} 
\label{dtdef}
d_N & = & \frac{1}{T}
\int^T_{0} dt \, \exp ( i\Omega t) 
\int d^3 {\bf r} \, \Psi_f ({\bf r}, \, t)^* \, \hat{d}_{\epsilon} \, 
\Psi_i ({\bf r}, \, t) ~.
\end{eqnarray} 
Here $\hat{d}_{\epsilon} = {\mbox {\boldmath $\epsilon$}}  \cdot {\bf r}$
is an operator of the dipole moment (the atomic units are used),
$\Psi_i$ and $\Psi_f$ are the initial and final states of 
the atomic system dressed by the laser field with
the frequency $\omega = 2\pi /T$, $\Omega = N \omega$. 

Both experiment and theory concentrate almost exclusively 
on the case when the initial and final states coincide. 
One can employ the exact time-dependent Green function 
$G(t, t^\prime)$ to construct 
the field-dressed states developed out of the initial 
(field-free) stationary state $\Phi_a$
\begin{eqnarray}\label{Psia}
\Phi_a({\bf r}, t) = \varphi_a({\bf r}) \, \exp( - i E_a t) ~,
\quad \quad
H_a \, \varphi_a = E_a \, \varphi_a ~ ,
\end{eqnarray}
where $H_a = \frac{1}{2}{\bf p}^2 + V_a({\bf r})$ is the 
atomic system Hamiltonian, $V_a({\bf r})$ is the interaction 
of the active electron with the core. 
This results in the expression  
(cf. Becker {\it et al}\/ 1997)
\begin{eqnarray} \label{df}
d_N & = &
\frac{1}{T}\int_0^T dt
\int^t_{-\infty} d t^\prime 
\langle \Phi_a(t) \mid \exp (i \Omega t) \, 
\hat{d}_\epsilon \, G(t, t^\prime) \,
V_F(t^\prime) \mid \Phi_a(t^\prime) \rangle ~,
\end{eqnarray}
where 
$V_F({\bf r}, t) = {\bf F} \cdot {\bf r} \, \cos \omega t$
is the interaction between the active electron and the laser 
wave with the electric field amplitude strength ${\bf F}$
in the dipole-length gauge.
Eq.(\ref{df}) implies that the high harmonic $\Omega$
is emitted {\it after}\/ the absorption of several low-frequency
laser quanta, $t \ge t^\prime $. Strictly speaking, there are 
'time-reversed' processes in which the radiation
of the high harmonic {\it precedes}\/ the absorption of some
laser quanta. However, for a large number of the quanta absorbed 
such a mechanism is suppressed and is therefore omitted in 
(\ref{df}) together with the continuum-continuum transitions 
(the latter approximation is a rather standard one, see, e.g., 
Lewenstein {\it et al}\/ (1994)).

The next basic approximation is to discard the effect of the 
atomic core potential $V_a$ on the Green function $G$
that allows one to represent it via the standard Volkov 
wave functions $\Phi_{\bf q}({\bf r}, t)$
\begin{equation}\label{grfu}
G({\bf r}, t; {\bf r}^\prime, t^\prime)=
-i \int  \frac{d^3 {\bf q}}{(2\pi)^3}    
\Phi_{\bf q}({\bf r}, t) \,
\Phi_{\bf q}^*({\bf r}^\prime, t^\prime), \quad t>t^\prime .
\end{equation}
Similar assumption underlies the Keldysh (1964)
model, whose recent {\it adiabatic}\/ modification 
(Gribakin and Kuchiev 1997a,b, Kuchiev and Ostrovsky 1998) 
gives fully reliable quantitative 
results for photodetachment. A useful extension of the Keldysh
model accounts for the Coulomb electron-core interaction 
(Ammosov {\it et al}\/ 1986, Krainov 1997).

Generally the correct description of the high Fourier components 
$d_N$ represents a formidable theoretical task. 
Its numerical implementation via solving the non-stati\-onary
Schr\"{o}dinger equation requires both a supercomputer and exceptional 
effort. In the representation (\ref{df}) the difficulty lies
in the strong variation of the integrand as a function of the 
time variables $t$, $t^\prime$. The crucial simplification
is gained by using the {\it factorization technique}\/
Kuchiev (1995) which allows us to disentangle the integration 
variables at a price of introducing an extra summation; 
very importantly, this summation is physically meaningful as
it corresponds to the contributions of different ATI channels. 
The integration over the intermediate momenta ${\bf q}$ 
[coming from (\ref{grfu})] is carried out in closed form. 
Some minor additional approximations [see, for instance, 
Eq.(\ref{R}) below; the detailed derivation is to be published
elsewhere] brings us to the appealing representation: 
\begin{eqnarray} \label{dsum}
d_N & = & \sum_{m=-\infty}^\infty \, 
\sum_{\sigma = \pm1} d_{Nm}^{(\sigma)} ~, 
\\ \label{Mpref}
d_{N \, m}^{(\sigma)} & = & - \,
\int \limits_{0}^{T} dt 
\int \limits_{0}^{T}dt^\prime \:
\langle \Phi_a(t) \mid \exp ( i \Omega t ) \, \hat{d}_\epsilon \mid 
\Phi_{{\bf K}_m} (t) \rangle
\nonumber \times \\ 
& \times & 
\frac{1}{ 2 \pi T^2R(t,t^\prime) }
\langle \Phi_{{\bf K}_m} (t^\prime) \mid V_F(t^\prime) \mid
\Phi_a (t^\prime) \rangle ~.
\end{eqnarray}
Here the vector ${\bf K}_m$ has an absolute value
\begin{equation}\label{Kmnm}
K_{m} =\left[ 2 \left ( m \omega -
F^2 / (4 \omega^2) + E_a\right ) \right]^{1/2}
\end{equation} 
and is parallel or antiparallel to ${\bf F}$
\begin{equation} \label{Ksigma}
{\bf K}_m = \sigma K_m \, {\bf F}/F, \quad ~ \sigma = \pm 1 ~. 
\end{equation}
The physical interpretation of (\ref{dsum}), (\ref{Mpref}) is 
based on the observation that the amplitude of $m$-photon 
detachment of the initial state $\Phi_a$ within 
the Keldysh model is 
\begin{eqnarray} \label{AKeld}
A_m({\bf p}) = \frac{1}{T} \, \int \limits_{0}^{T} dt^\prime \,
\langle \Phi_{{\bf p}} (t^\prime) \mid V_F(t^\prime) \mid 
\Phi_a (t^\prime) \rangle ~ .
\end{eqnarray}
In the right hand side of (\ref{AKeld}) the index $m$ is implicit.
It enters via the absolute value of the final electron momentum 
${\bf p}$ which is subject to the energy conservation constraint.
Namely, the absolute value of ${\bf p}$ is given by the right 
hand side of the formula (\ref{Kmnm}),
where 
$F^2/(4 \omega^2)$ is the electron quiver energy 
in the laser field. This shows  that the vector ${\bf K}_m$ 
entering the representation (\ref{Mpref}) of HG amplitude 
component is exactly the physical electron translational
momentum in the $m$-th ATI channel, but with the specific 
directions (\ref{Ksigma}).

  From the Volkov state $\Phi_{{\bf p}}$ the electron can recombine 
back to the bound state $\Phi_a$ with emission of the photon 
of frequency $\Omega$. This process with the amplitude
\begin{eqnarray} \label{rec}
C_{Nm}({\bf p}) = - \frac{1}{2 \pi T} \int \limits_{0}^{T} dt \,
\langle \Phi_a(t) \mid \exp ( i \Omega t ) \, 
\hat{d}_\epsilon \mid \Phi_{{\bf p}} (t) \rangle 
\end{eqnarray}
is possible only in the presence of the laser field from which
the necessary  $N-m$ low-frequency quanta $\omega$ are 
gained, that justifies its name {\it stimulated recombination}.

One readily notices that the integrand in (\ref{Mpref}) 
bears a striking resemblance to the product of the integrands 
in (\ref{AKeld}) and (\ref{rec}). However, the complete separation 
of integrations in $t$ and $t^\prime$ variables is prevented 
by the factor $1/R(t, t^\prime)$ stemming from the chain 
of equations
\begin{eqnarray} \label{R}
{\bf R}({\bf r},{\bf r}^\prime;t,t^\prime)
& = 
({\bf F}/\omega^2) \left( \cos \omega t - 
\cos \omega t^\prime \right) + {\bf r} - {\bf r}^\prime \approx
\nonumber \\ 
& \approx  {\bf R}(t, t^\prime) \equiv
({\bf F}/\omega^2) \left( \cos \omega t - \cos \omega t^\prime \right) .
\end{eqnarray}
The latter approximation is applicable provided $F/\omega^2$ exceeds 
the typical dimensions of localization of the active electron in 
the initial state $\Phi_a$, that holds in most practical 
situations. Classically $R(t, t^\prime)$ gives the distance between 
the electron positions at the moments $t$ and $t^\prime$ 
due to electron wiggling in the laser field.
$1/R$ could be named an {\it expansion factor}\/
since in the absence of the laser field 
it describes conventional decrease of the amplitude in a
spherical wave as it expands in 3D space. When the laser 
field is operative, the form of the expansion factor is 
drastically modified according to Eq.(\ref{R}).
Hence the interpretation of the expression (\ref{Mpref}) 
is that the electron first is transferred to the $m$-th ATI channel, 
then propagates in space under the influence of the 
laser wave and finally recombines to the initial state emitting
the photon with the frequency $\Omega$.
The contributions of all paths labeled by $m$ add up 
coherently as shown by Eq.(\ref{dsum}).

  Following the {\it factorization technique} (Kuchiev 1995) we simplify 
Eq.(\ref{Mpref}) further on by performing the $t'$ integration 
by the saddle point method. This is justified since the integrand 
in (\ref{Mpref}) contains a large phase factor
$\exp \left[ i S(t^\prime) \right]$ with the classical action $S(t)$
\begin{eqnarray}
S(t^\prime) = \frac{1}{2} \int^{t^\prime} d \tau \left[{\bf p} + 
({\bf F}/\omega) \sin \omega \tau \right]^2
-E_a t^\prime ~.
\end {eqnarray}
The saddle point positions $t^\prime=t_{m \mu}^\prime$ in the 
complex $t^\prime$ plane are defined by the equation 
\begin{eqnarray} \label{seqa}
S^\prime(t_{m \mu}^\prime) = 0 ~.
\end{eqnarray}
Similar {\it adiabatic}\/ approximation in a simpler case
of photodetachment casts the ATI amplitude (\ref{AKeld}) as
\begin{eqnarray} 
\label{A}
A_m ({\bf p}) & = & \sum_{\mu = 1,2} \, 
A_{m \, \mu} ({\bf p}) ~, 
\\ 
\label{Amu}
A_{m \, \mu} ({\bf p}) 
& = & - \frac{1}{\omega} \, A_a \, 
Y_{lm}(\hat{{\bf p}}) \,
\sqrt{\frac{2 \pi i}{ S^{\prime \prime}_\mu}} \,
\exp \left( i S_\mu \right) ~.
\end{eqnarray}
In the plane of the complex-valued time the saddle points $t_{m \mu}$ lie
symmetrically with respect to the real axis. There are four saddle 
points in the interval $ 0 \leq {\rm Re} \, t_{m \mu} \leq T$, two
of them lying in the upper half plane ${\rm Im} \, t_{m \mu} > 0$.
Only these two saddle points are operative in the contour
integration being included into the summation over $\mu = 1,2$.  
In (\ref{Amu}) $l$ is the active electron orbital momentum 
in the initial state, $\kappa = \sqrt{ - 2 E_a}$, 
$A_a$ is the coefficient in the wave function $\phi_a({\bf r})$
asymptote. For more details see the papers 
by Gribakin and Kuchiev (1997a,b),
where the approximation based on (\ref{A})-(\ref{Amu}) was 
demonstrated to be very efficient and accurate for multiphoton 
detachment.

After carrying out the $t^\prime$ integration in (\ref{M}) by the
saddle point method we arrive to our major result
\begin{eqnarray} \label{dfin}
d_N = 2 \, \, \sum_m \, A_{m \, \mu} ({\bf K_m}) \, 
B_{N m \, \mu}({\bf K}_m) ~, 
\end{eqnarray}
where the factor
\begin{eqnarray}
\label{M}
B_{N m \, \mu}({\bf K}_m) &=& - \frac{1}{2 \pi T}  \times 
\\ \nonumber 
& \times & \, \int \limits_{0}^{T} dt \,
\frac{
\langle \Phi_a(t) \mid \exp ( i \Omega t ) \, 
\hat{d}_\epsilon \mid \Phi_{{\bf K}_m} (t) \rangle
}
{(F/\omega^2)(\cos \omega t^\prime_{m \, \mu} - \cos \omega t )} 
\end{eqnarray}
describes jointly the 3D-wave expansion and SR. These two effects
could be further factorized using the approximation
$|\cos \omega t^\prime_{m \, \mu}| \gg |\cos \omega t |$  
(Kuchiev 1995): 
\begin{eqnarray} \label{Ma}
B_{N m \, \mu}({\bf K}_m) =  \frac{1}{R_{m \mu}} \, C_{Nm}({\bf K}_m) ~,
\end{eqnarray}
where $1/R_{m \mu} = \omega^2 /(F \cos t_{m \mu}^\prime)$ 
is the laser-modified expansion factor in its simplest form.  

Now it is worthwhile to comment more on the physics of 
HG as implemented in Eq.(\ref{dfin}). On the first stage
of the three-step process the electron absorbs $m$ laser
photons with the amplitude $A_{m \, \mu} $.
In order to contribute to HG the photoelectron has to 
return to the parent atomic core where SR is solely 
possible. The amplitude of return is described by the expansion
factor $1/R$. At the third step the electron collides
with the core virtually absorbing $N-m$ photons from 
the laser field and emitting the single high-frequency 
quantum $\Omega$ as it recombines to the bound state. 
This SR process has the amplitude $C_{Nm}$. 
The summation over $m$ in the total amplitude $d_N$ 
(\ref{dfin}) takes into account interference
of the transitions via different intermediate ATI 
channels.

The nontrivial point is the probability for the ATI electron 
to return to the core. Intuitively, one could anticipate that 
such a process is suppressed, because the most {\em natural}\/ 
behavior for the electron would be simply to leave the atom. 
The proper description of the suppression plays substantial 
role in the theory. According to the physical image of the ATI 
process worked out in the adiabatic approach 
(Gribakin and Kuchiev1997a,b), 
after tunneling through the time-dependent barrier
the ATI electron emerges from under the barrier  at some 
point which is well separated from the core.
As a result this point becomes the source of an expanding 
spherical wave. This occurs twice per each cycle of the 
laser field, at the two moments of time $t_{m \mu}^\prime$ 
when the source-points lie up and down the field ${\bf F}$ 
from the core. The interference of the two spherical waves
originating from the two different source-points results 
in non-trivial patterns in the angular ATI photoelectron
distributions obtained from (\ref{A})-(\ref{Amu})  
(Gribakin and Kuchiev 1997a,b, Kuchiev and Ostrovsky 1998) 
in agreement with the available theoretical and experimental data.
The probability for the ATI electron to return to the core  
from the source-point is governed by the expansion factor $1/R$ 
and by the direction of propagation. At each of the moments 
$t_{m \mu}^\prime$ only {\it one}\/ of the two possible 
directions of ${\bf K}_m$, labeled in (\ref{dsum}) by 
$\sigma = \pm 1$, results in the electron eventually
approaching the core. For the opposite direction of ${\bf K}_m$ 
the electron recedes from the core and does not come back. 
In other words, for each direction of ${\bf K}_m$ only one of the
two saddle points $t_{m \mu}^\prime$ contributes to HG.
Since both values of $\sigma$ give identical contributions, 
summation over $\sigma$ simply gives an extra factor of 2 
in (\ref{dfin}).

The practical calculations of $B_{N m \, \mu}$ or 
$C_{Nm}$ can be conveniently performed using the Fourier
transform of the bound state wave function $\phi_a$.
After that one has to carry out a single numerical
integration over the finite interval of time $t$, or, 
alternatively, to resort to the saddle-point method that
provides purely analytical formulae.

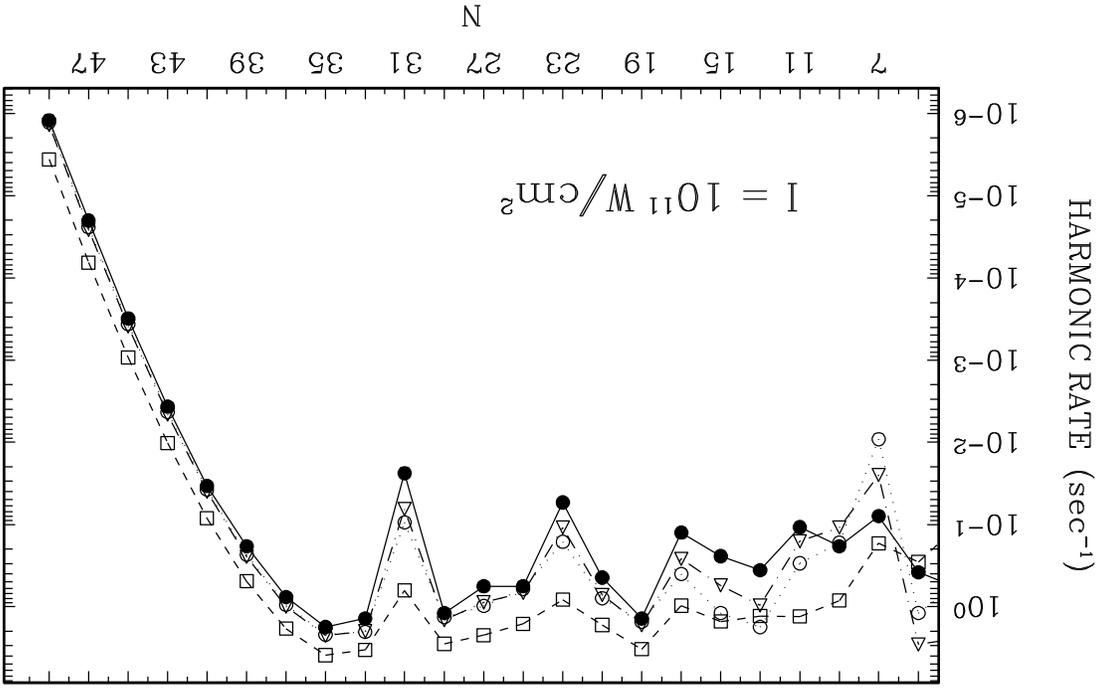
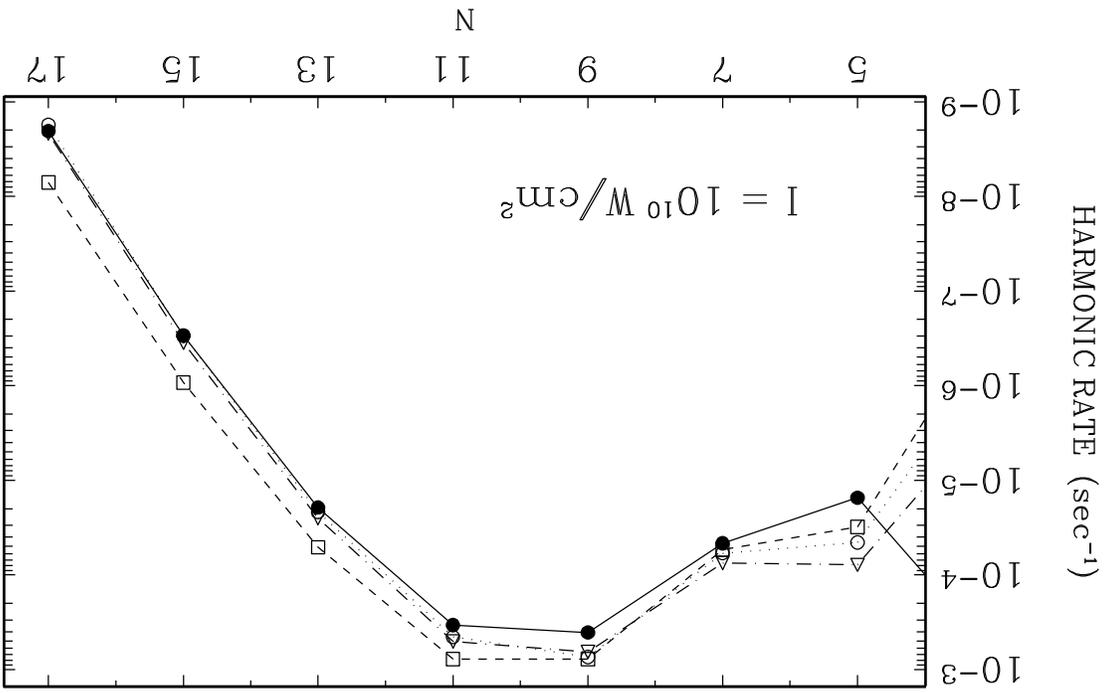
\begin{figure}
\caption{ \label{Fig1}
Harmonic generation rates 
(\protect \ref{RN})
(in sec$^{-1}$) for H$^-$ ion 
in the laser field with the frequency $\omega = 0.0043$ 
and various values of intensity $I$ as indicated in the plots.
Closed circles - results obtained by Becker {\it et al}\/
(1994), open circles - present calculations
in the dipole-length gauge
using the expression (\protect \ref{M}) for $B_{N m \, \mu}$,
open squares - same but with the simplified formula
(\protect \ref{Ma}) for $B_{N m \, \mu}$;
open triangles - present calculations in the dipole-velocity gauge. 
The symbols are joined by lines to help the eye.
}
\end{figure}

Based on physical arguments, we extend the summation in (\ref{dfin}) 
only over open ATI channels with the real values of $K_m$.
We present the rates of generating the $N$-th harmonic radiation
\begin{eqnarray} \label{RN}
{\cal R}_N \equiv \frac{\omega^3 N^3}{2 \pi c^3} \, \mid d_N \mid^2 
\end{eqnarray}
as introduced by Becker {\it et al}\/ (1994) (and denoted by these
authors as $d R_N /d\Omega_{\bf K}$); $c$ is the velocity of light.
Some typical results for the HG by H$^-$ ion in the
$\omega = 0.0043$ laser field are shown 
in Fig. \ref{Fig1} for the smallest and largest field 
intensities considered in the paper by Becker {\it et al}\/ (1994).  
We take the binding energy corresponding to 
that of H$^-$ ($\kappa = 0.2354$), but
replace the true value $A_a=0.75$ (Radzig and Smirnov 1985) 
by unity since 
this corresponds to the zero-range potential model 
used by Becker {\it et al}\/ (1994).
For the real H$^-$ ion the results shown in Fig.\ref{Fig1} are
to be scaled by a factor $A_a^4 N_e$, where
$N_e=2$ accounts for the presence of two active electrons
in H$^-$. 

The HG spectrum is known to consist generally of the initial
rapid decrease, the plateau domain and the rapid fall-off region.
The present theory is designed to describe the high
HG but not the initial decrease
(which in the case considered is noticeable only for 
one or two lowest harmonics).
On the large-$N$ side our results employing Eq.(\ref{M}) 
coincide with those obtained by Becker {\it et al}\/ (1994) 
within the plot scale. The deviations increase 
as $N$ decreases, albeit remarkably the structures
in $N$-dependence of the rates are well reproduced.
The approximation (\ref{Ma}) somewhat overestimates
HG rate, but still retains the structure, though smoothed.

In the summation over ATI channels (i.e. over $m$) the coherence
is very important, since the large number of terms is comparable
in modulus, but have rapidly varying phases.
Many ATI channels contribute to HG for each $N$
(contrary to tentative conclusion by Eberly {\it et al}\/ (1989)).

Although the length gauge is known to be superior for the description 
of ATI within the adiabatic approximation 
(Gribakin and Kuchiev 1997a), the situation
is not that straightforward for the high-energy photon.
Therefore our calculations were reiterated with replacement
of the $\hat{d}_\epsilon$ operator in (\ref{M}) by its 
dipole-velocity counterpart. The agreement between the two 
forms proves to be very good, see Fig. \ref{Fig1}.

As a summary, the three-step mechanism of the harmonic
generation is ultimately justified. It is implemented in fully	
quantum relations expressing its amplitude via amplitudes
of the above-threshold ionization and stimulated recombination.
The theory is quantitatively reliable and easy to apply.
It gives an important physical insight being a particular
realization of the general {\it atomic antenna}\/ mechanism.

The authors are thankful to the Australian Research Council for 
the support. V.~N.~O. acknowledges the hospitality of the staff of 
the School of Physics of UNSW where this work has been
carried out. The stimulating discussions with G.~F.~Gribakin
are appreciated.

\section*{References}
\begin{harvard}


\item[]
Ammosov M V, Delone N B and Krainov V P 1986
{\it Zh. Eksp. Teor. Fiz.}\/ {\bf 91} 2008
[1986 {\it Sov. Phys.-JETP}\/ {\bf 64} 1191]

\item[]
Becker W, Long S and McIver J K 1994 {\it Phys. Rev. A}\/ 
{\bf 50} 1540 

\item[]
Becker W, Lohr A, Kleber M and Lewenstein M 1997 
{\it Phys. Rev. A}\/ {\bf 56} 645 

\item[]
Corkum P B 1993 {\it Phys. Rev. Lett.}\/ {\bf 71} 1994 

\item[]
Eberly J H, Su Q and Javanainen J 1989 {\it Phys. Rev. Lett.}\/ 
{\bf 62} 881 

\item[]
Gribakin G F and Kuchiev M Yu 1997a {\it Phys. Rev. A}\/ {\bf 55} 3760 

\item
\dash 1997b {\it J. Phys. B: At. Mol. Opt. Phys.} {\bf 30} L657 
(Corrigendum: 1998 {\it J. Phys. B: At. Mol. Opt. Phys.} {\bf 31} 3087)

\item[]
L.'Huillier A, Schafer K J, and Kulander K C 1991
{\it Phys. Rev. Lett.} {\bf 66}, 2200 

\item[]
Keldysh L V 1964 {\it Zh. Eksp. Teor. Fiz.}\/ {\bf 47} 1945 
[1965 {\it Sov. Phys.-JETP}\/ {\bf 20} 1307]

\item[]
Krainov V P 1997 {\it J. Opt. Soc. Am. B}\/ {\bf 14} 425 

\item[]
Kuchiev M Yu 1987 {\it Pis'ma Zh. Eksp. Teor. Fiz.}\/ {\bf 45} 319
[1987 {\it JETP Letters}\/ {\bf 45} 404]

\item[]
\dash 1995 {\it J. Phys. B: At. Mol. Opt. Phys.}\/ {\bf 28} 5093

\item[]
\dash 1996 {\it Phys. Lett. A}\/ {\bf 212} 77

\item[]
Kuchiev M Yu and Ostrovsky V N 1998 
{\it J. Phys. B: At. Mol. Opt. Phys.}\/ {\bf 31} 2525

\item[]
Lewenstein M, Balcou Ph, Ivanov M Yu, L'Huillier A and Corkum P B
1994 {\it Phys. Rev. A}\/ {\bf 49} 2117

\item[]
Macklin J J, Kmetec J D, and Gordon C L 1993
{\it Phys. Rev. Lett.}\/ {\bf 70} 766

\item[]
Radzig A A and Smirnov B M 1985 {\it Reference Data on Atoms,
Molecules and Ions} (Berlin: Springer)

\item[]
Schafer K J , Yang B, DiMauro L F and Kulander K C 1993
{\it Phys. Rev. Lett.}\/ {\bf 70} 1599

\item[]
Zaretskii D F and Nersesov E A 1996 
{\it Zh. Eksp. Teor. Fiz.}\/ {\bf 109} 1994 
[1996 {\it JETP}\/ {\bf 82} 1073]


%

\end{harvard}

\Figures

\end{document}